\documentclass[%
 aip,
 amsmath,amssymb,
 reprint,%
]{revtex4-1}

\usepackage{graphicx}
\usepackage{dcolumn}
\usepackage{bm}

\usepackage[utf8]{inputenc}
\usepackage[T1]{fontenc}
\usepackage{mathptmx}
\usepackage{etoolbox}

\makeatletter
\def\@email#1#2{%
 \endgroup
 \patchcmd{\titleblock@produce}
  {\frontmatter@RRAPformat}
  {\frontmatter@RRAPformat{\produce@RRAP{*#1\href{mailto:#2}{#2}}}\frontmatter@RRAPformat}
  {}{}
}%
\makeatother
\begin{document}

\title[]{Spectroscopic investigations on trivalent ruthenium ions in ruthenium perovskite oxide thin films}
\author{S. Nakata}
\email{nakata@sci.u-hyogo.ac.jp}
\affiliation{Department of Material Science, Graduate School of Science, University of Hyogo, Ako, Hyogo 678-1297, Japan}

\author{R. Takahashi}
\affiliation{Department of Material Science, Graduate School of Science, University of Hyogo, Ako, Hyogo 678-1297, Japan}

\author{R. Matsumoto}
\affiliation{Department of Material Science, Graduate School of Science, University of Hyogo, Ako, Hyogo 678-1297, Japan}

\author{L.-F. Zhang}
\affiliation{Department of Applied Physics and Quantum Phase Electronics Center, University of Tokyo, Tokyo 113-8656, Japan}

\author{H. Sumida}
\affiliation{Mazda Motor Corporation, 3-1, Shinchi, Fuchu, Aki, Hiroshima 730-8670, Japan}

\author{S. Suzuki}
\affiliation{Laboratory of Advanced Science and Technology for Industry, University of Hyogo, 3-1-2, Koto, Kamigori, Ako, Hyogo 678-1205, Japan}

\author{T. C. Fujita}
\affiliation{Department of Applied Physics and Quantum Phase Electronics Center, University of Tokyo, Tokyo 113-8656, Japan}

\author{M. Kawasaki}
\affiliation{Department of Applied Physics and Quantum Phase Electronics Center, University of Tokyo, Tokyo 113-8656, Japan}
\affiliation{RIKEN Center for Emergent Matter Science (CEMS), Wako 351-0198, Japan}

\author{H. Wadati}
\affiliation{Department of Material Science, Graduate School of Science, University of Hyogo, Ako, Hyogo 678-1297, Japan}
\affiliation{Institute of Laser Engineering, Osaka University, Suita, Osaka 565-0871, Japan}

\date{\today}

\begin{abstract}
The $d^5$ electron configurations under the crystal field, spin-orbit coupling, and Coulomb interaction give rise to a plethora of profound ground states. Ruthenium perovskite oxides exhibit a number of unconventional properties yet the Ru$^{4+}$ state ($4d^4$) is usually stable in these materials. In this regard, Ru$^{3+}$ ions in perovskite materials are expected to be a mesmerising playground of $4d^5$ electron configurations. Here, we report measurements of x-ray photoemission spectroscopy on recently synthesized perovskite ruthenium oxide thin films, LaRuO$_3$ and NdRuO$_3$, whose valence state of the ruthenium ions is trivalent. We discuss correlation and spin-orbit effects from the valence-band spectra, in particular  an additional peak structure around 3-5 eV, reminiscent of the so-called 3 eV peak observed in Sr$_2$RuO$_4$. Moreover, we find that the core-level spectra of these materials are quantitatively different from those in other ruthenates which possess Ru$^{4+}$  ions, e.g., SrRuO$_3$. We therefore argue that the core level spectra of LaRuO$_3$ and NdRuO$_3$ are peculiar to the  Ru$^{3+}$ states. 
\end{abstract}

\maketitle


The interplay between the spin-orbit interaction and Coulomb interaction has been studied as one of central topics in modern condensed matter physics \cite{doi:10.1146/annurev-conmatphys-020911-125138,doi:10.1146/annurev-conmatphys-020911-125045}.  
Notably, the physical properties of transition-metal compounds strongly depend on the number of electrons in the $d$ orbitals \cite{PhysRevLett.107.256401}.
In particular, the $d^5$ electron configuration under the crystal field of the honeycomb lattice structure, which gives rise to the filled $j=3/2$ quartet and half-filled $j=1/2$ doublet manifolds due to the spin-orbit entanglement between the $t_{2g}$ orbitals and electron spin $1/2$, has been attracted in part because the ground state of these materials have theoretically been proposed as a Kitaev spin liquid state \cite{PhysRevLett.102.017205} and experimental evidence to support this proposal have recently been accumulated in several candidate materials, e.g., $\alpha$-RuCl$_3$ \cite{PhysRevB.90.041112} and $\beta$-Li$_2$IrO$_3$ \cite{PhysRevLett.114.077202}.

Alongside the research avenue on the spin liquid candidate materials, the $d^5$ system in the perovskite structure is another mesmerizing playground with the spin-orbit interaction. For instance, Sr$_2$IrO$_4$ is understood to be a Mott insulator owing to the Mott-Hubbard splitting of the $j=1/2$ doublet energy level although one would naively expect a metallic state because of the spatially extended 5$d$ orbitals resulting in the large $t_{2g}$ bandwidth without considering the spin-orbit interaction \cite{PhysRevLett.101.076402}. In the 4$d$ element series, the $4d^5$ electronic configuration is chemically stable in rhodium compounds, e.g., SrRhO$_3$ \cite{PhysRevB.64.224424}, Sr$_2$RhO$_4$ \cite{Perry_2006}, and Sr$_3$Rh$_2$O$_7$ \cite{PhysRevB.66.134431}, yet these materials are paramagnetic metals. A number of ruthenium perovskite compounds including a ferromagnetic metal SrRuO$_3$, an unconventional superconductor Sr$_2$RuO$_4$ \cite{Maeno1994Supercondu}, a Mott insulator Ca$_2$RuO$_4$ \cite{doi:10.1143/JPSJ.66.1868} exhibit profound properties and therefore their electronic structures have extensively been studied. Nevertheless, the valence number of ruthenium ions in these materials is typically four and trivalent ruthenium ions are usually unstable in a solid. As a consequence, the 4$d^5$ analogues in perovskite ruthenium oxides have rarely been studied. Recently, Zhang {\it et al.} reported the fabrication of epitaxially stabilized perovskite ruthenium oxide thin films: LaRuO$_3$ and NdRuO$_3$ \cite{2305.09201}. Remarkably, Ru $K$-edge x-ray absorption spectroscopy (XAS) revealed that the valence state of ruthenium ions in these compounds is trivalent (4$d^5$) despite chemical instabilities. These high-quality thin films also exhibited the substantially better conductivity in electrical transport measurements than powder samples previously reported. 

\begin{figure*}
  \centering
  \includegraphics[angle = 0, width = 0.95\textwidth, clip=true]{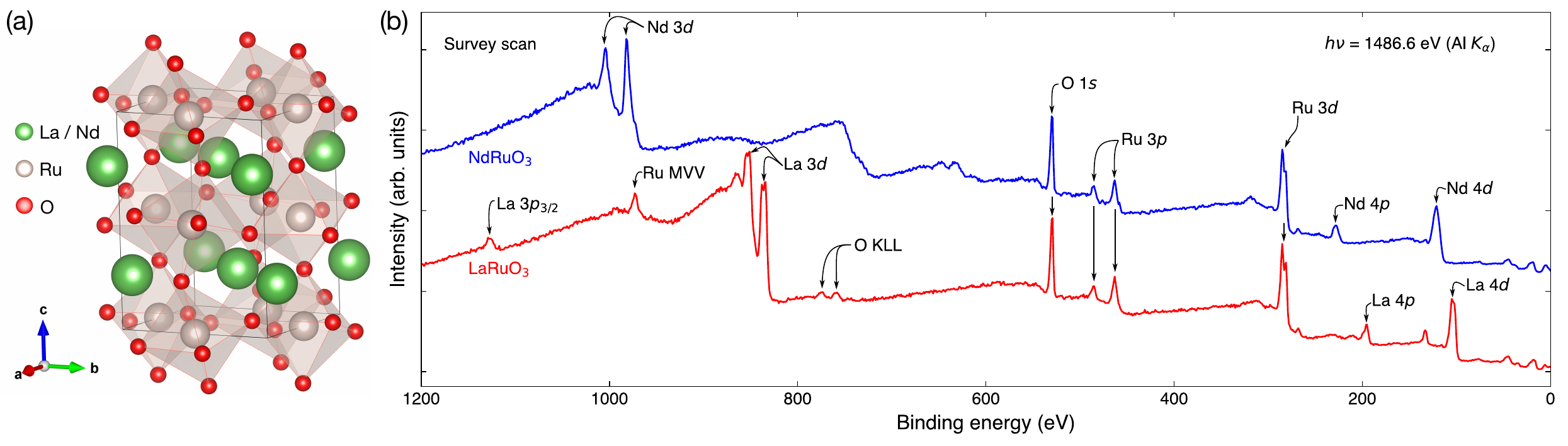}
  \caption{(a) Crystal structure of $Ln$RuO$_3$ ($Ln$ = Nd, La) \cite{Momma:db5098}. (b) Survey SXPES scans of the NdRuO$_3$ and LaRuO$_3$ thin films collected at room temperature. \label{fig:1}}
\end{figure*}

In contrast to other 4$d$ elements such as molybdenum whose different valence states yield distinct core-level photoemission spectra \cite{PhysRevB.90.205131}, little was known other than Ru$^{4+}$ ions in the case of ruthenates. Ru 3$d$ core-level spectra yield a wealth of information of electron correlation effects in terms of well-screened and poorly-screend states \cite{PhysRevB.72.060404,PhysRevLett.93.126404}. To this end, we have performed soft x-ray photoemission spectroscopy (SXPES) and hard x-ray photoemission spectroscopy (HAXPES) measurements on LaRuO$_3$ and NdRuO$_3$ thin films. The good conductivity of these new ruthenate thin films enables us to scrutinize the trivalent state of ruthenium ions in the perovskite oxides by means of photoemission experiments without ambiguities of charging effects. We find that all the ruthenium and oxygen core-level spectra in both compounds are similar each other but clearly distinct from those of a SrRuO$_3$ thin film (4$d^4$) \cite{PhysRevB.56.6380}. We thus attribute these observations in LaRuO$_3$ and NdRuO$_3$ to the trivalent state of ruthenium ions. The core-level spectra of cations (La and Nd) also support this argument. Moreover, we find an additional peak structure near the Fermi level, reminiscent of the well known 3 eV peak observed in Sr$_2$RuO$_4$ whose origin has been debated for decades \cite{PhysRevB.53.8151,PhysRevB.75.035122,PhysRevB.77.046102,PhysRevB.77.046101,PhysRevB.93.075125,PhysRevB.70.153106}.


Epitaxial ruthenium oxide thin films (LaRuO$_3$ and NdRuO$_3$) were fabricated on SrTiO$_3$ (001) substrates by means of the pulsed laser deposition technique. Details for the sample growth and subsequent annealing processes can be seen in the recent report by Zhang {\it et al.} \cite{2305.09201}.
The crystal structure of both compounds was confirmed to be a perovskite structure (Fig. \ref{fig:1} (a)) by x-ray diffraction. 
The in-plane sample dimensions are 5 $\times$ 5 mm$^2$. The film thickness of LaRuO$_3$ and NdRuO$_3$ was controlled to be approximately 25 nm and 14 nm, respectively.

SXPES measurements on the LaRuO$_3$ and NdRuO$_3$ films were carried out using a spectrometer with a monochromatic Al $K_\alpha$ source ($h\nu$ = 1486.6 eV) implemented in PHI 5000 VersaProbe system (ULVAC-PHI Inc.).
The combined instrumental energy resolution was set to be $\Delta E \sim$ 450 meV. The angle of the photoelectron trajectory is normal to the sample surface. HAXPES measurements on the LaRuO$_3$ and NdRuO$_3$ films were carried out at BL24XU of SPring-8. The incidence angle of the horizontally linearly polarized hard x-ray ($h\nu$ = 7994 eV) was set at 2$^\circ$. The photoemission spectra were collected with the Scienta R-4000 electron energy analyzer. The combined energy resolution is around 270 meV. All SXPES and HAXPES spectra were collected at room temperature. The binding energy of the spectra was calibrated using the Fermi edge of silver in electrically contact with samples.


\begin{figure}
  \centering
  \includegraphics[angle = 0, width = 0.48\textwidth, clip=true]{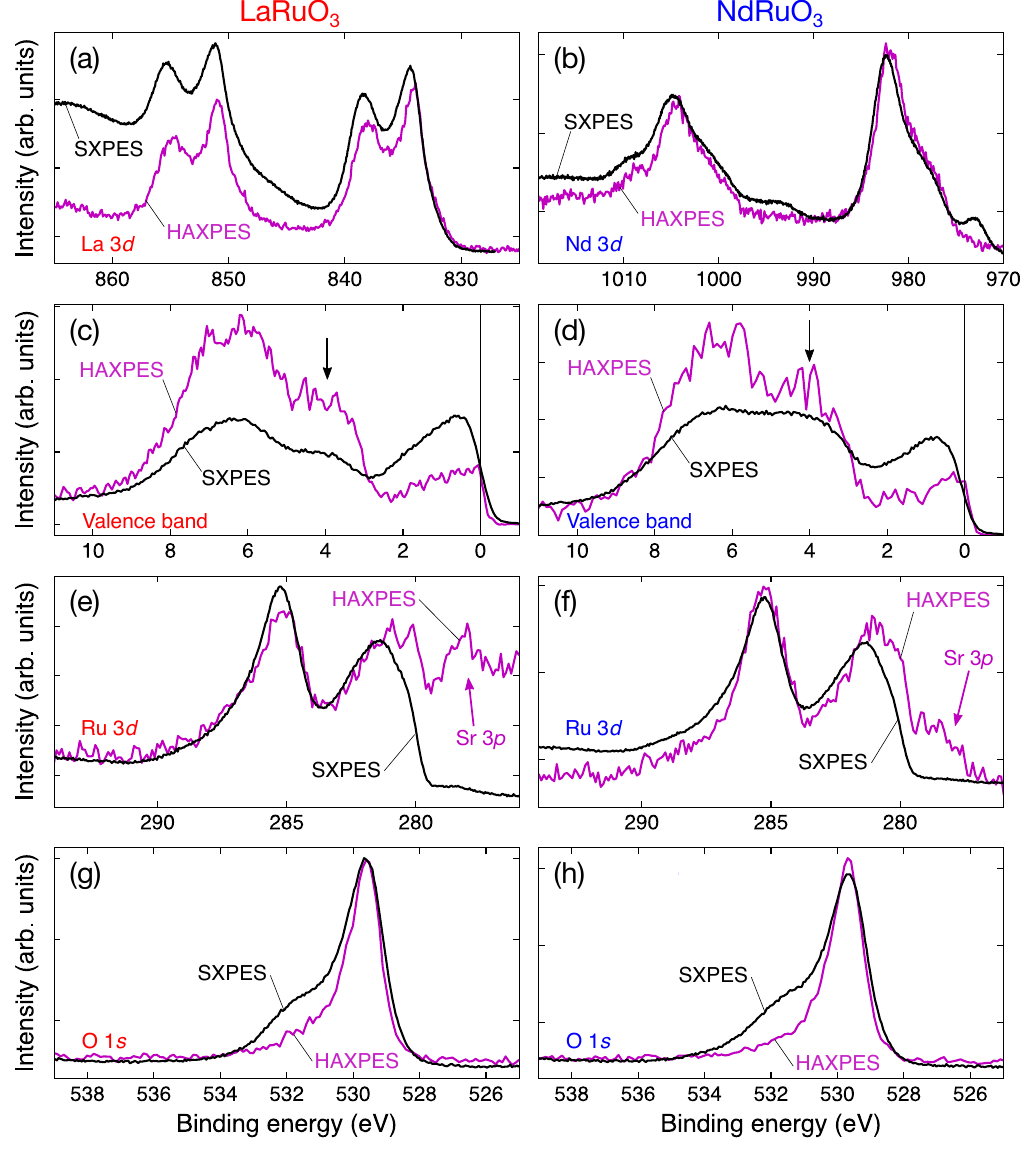}
  \caption{Core-level and valence band spectra of NdRuO$_3$ and LaRuO$_3$ collected with SXPES (black) and HAXPES (purple). (a) La 3$d$. (b) Nd 3$d$. (c,d) Valence band. (e,f) Ru 3$d$. (g,h) O 1$s$. For clarity, the spectral intensities are normalized. \label{fig:4}}
\end{figure}

Figure \ref{fig:1} (b) displays survey scans of the LaRuO$_3$ and NdRuO$_3$ thin films. The core-level peaks expected from the nominal compositions are clearly recognizable. 
Before discussing the valence band and core-level spectra in detail, we first make a comparison between the SXPES and HAXPES data in Fig. \ref{fig:4}, as also done for SrRhO$_3$ thin films \cite{PhysRevB.101.085134}. The peak positions in all the SXPES and HAXPES spectra are nearly identical despite the significantly different electron inelastic mean free path of electrons excited by the soft and hard x-rays that we employed \cite{Hufner}. The spectral weight in the valence band spectra is yet different between SXPES and HAXPES (Fig. \ref{fig:4} (c,d)). The atomic photoemission cross section ratio of Ru 4$d$ $vs.$ O 2$p$ is 40 : 1 at 1486.6 eV and 70 : 1 at 8047.8 eV \cite{Yeh1985Atomic-sub}, therefore the different spectral weight in these two measurements is generally anticipated. However, this atomic picture is quantitatively inconsistent with our experiments if one merely associates the intensity near the Fermi energy and at 5-7 eV with the Ru 4$d$ and O $2p$ orbitals, respectively. This disagreement with the atomic picture indicates a hybridization between the Ru 4$d$ and O 2$p$ orbitals as in the case of other ruthenium compounds such as SrRuO$_3$ and Sr$_2$RuO$_4$ \cite{PhysRevB.93.075125}.
Moreover, the non-negligible intensity is discernible at 3-5 eV in both spectra. A similar feature is not clearly observed in SrRuO$_3$ but in Sr$_2$RuO$_4$ and this aspect will be further discussed later. In Fig. \ref{fig:4} (e, f), the HAXPES intensity below 280 eV is pronounced and could be attributed to Sr 3$p$ spectra of the SrTiO$_3$ substarates due to the longer inelastic mean free path. We ascribe a high energy shoulder in the O 1$s$ core level appearing in the SXPES spectra to surface contamination since this feature is substantially suppressed in bulk-sensitive HAXPES (Fig. \ref{fig:4} (g,h)). We therefore attribute the main peak at 529.7 eV to the oxygen ions in the ruthenate thin films in the following discussion.

\begin{figure}
  \centering
  \includegraphics[angle = 0, width = 0.44\textwidth, clip=true]{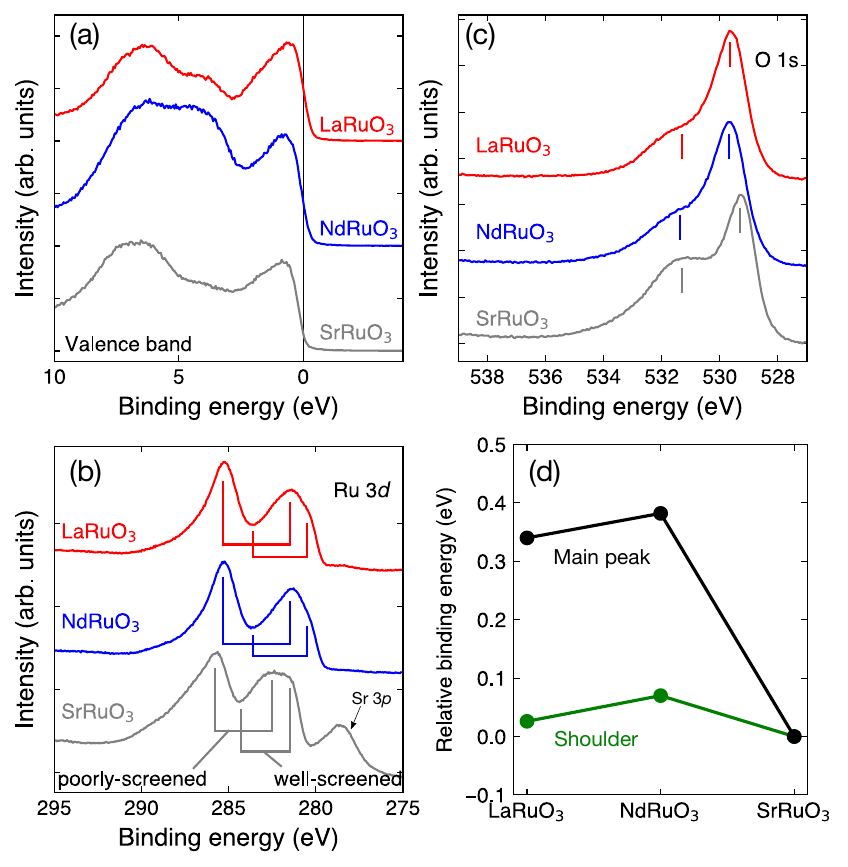}
  \caption{SXPES spectra of LaRuO$_3$ (red) and NdRuO$_3$ (blue). The spectra are vertically shifted for clarity. For comparison, the corresponding spectra of SrRuO$_3$ (gray) are also plotted. (a) Valence band. (b) Ru 3$d$. (c) O 1$s$. (d) The relative binding energy of the main peak and shoulder peak in the O 1$s$ spectra. \label{fig:2}}
\end{figure}

In light of these observations that our SXPES essentially measures bulk properties of the films, let us hereafter discuss the SXPES spectra of LaRuO$_3$ and NdRuO$_3$ in conjunction with an archetypical ruthenium oxide thin film SrRuO$_3$ whose rutenium valence number is four (Fig. \ref{fig:2}).

We first directly compare valence-band spectra (Fig. \ref{fig:2} (a)). The valence band spectra show the non-zero Fermi edge, indicative of the metallic states of these samples and consistent with transport measurements. The pseudospin splitting into the $j$=1/2 doublets and $j$=3/2 quartets was however not experimentally recognizable most likely because the energy resolution is not good enough to resolve these low energy features in the present work. In fact, the transition energy between these two spin states in another $d^5$ ruthenium compound $\alpha$-RuCl$_3$ investigated by resonant inelastic x-ray scattering (RIXS) was found to be approximately 250 meV \cite{Suzuki2021Proximate-} although more complicated band structures are anticipated in LaRuO$_3$ and NdRuO$_3$ in part because these materials are much more metallic than $\alpha$-RuCl$_3$. 

The binding energy of the Ru 3$d$ core levels of LaRuO$_3$ and NdRuO$_3$ are downwards shifted in comparison with SrRuO$_3$ (Fig. \ref{fig:2} (b)). By substituting a cation of SrRuO$_3$, such spectral shift was not observed \cite{PhysRevB.72.060404}. We therefore interpret that these shifts are not caused by the cation radius of La and Nd larger than Sr. We attribute this spectral shift to the difference between tetravalent and trivalent ruthenium valence states as also observed in tetravalent and sexivalent molybdenum compounds \cite{PhysRevB.90.205131}. A similar energy shift was reported for Ca$_2$RuO$_4$ thin films upon fluorination and interpreted as the transition from a tetravalent to trivalent valence state of ruthenium ions \cite{PhysRevMaterials.6.035002}.
Besides the energy shifts, the spectral shape of Ru 3$d_{5/2}$ provides spectroscopic insights regarding well-screened and poorly-screened features \cite{PhysRevLett.93.126404}. In LaRuO$_3$ and NdRuO$_3$, the poorly-screened feature is more pronounced than the well-screened feature compared to SrRuO$_3$, in which both features are comparable. Since it is known that the stronger electron correlation is the more (less) pronounced well-screened (poorly-screened) feature is, the electron correlation in LaRuO$_3$ and NdRuO$_3$ is expected to be weaker than SrRuO$_3$. This disparity could yield insight into the fact that SrRuO$_3$ exhibits the ferromagnetic ordering despite the absence of magnetic ordering in LaRuO$_3$ and NdRuO$_3$.

Compared to SrRuO$_3$ \cite{PhysRevB.86.235127}, another salient feature in the valence band spectra in both LaRuO$_3$ and NdRuO$_3$ is an additional peak structure around 3-5 eV (Fig. \ref{fig:2} (a)), which is reminiscent of the well known 3 eV peak of Sr$_2$RuO$_4$ whose origin has been discussed for decades \cite{PhysRevB.53.8151,PhysRevB.75.035122,PhysRevB.77.046102,PhysRevB.77.046101,PhysRevB.93.075125,PhysRevB.70.153106}. This feature has been associated with the lower Hubbard band of Ru 4$d$ states from SXPES and resonant SXPES measurements on the one 
hand \cite{PhysRevB.75.035122,PhysRevB.77.046102}, this is mainly ascribed to the oxygen states from the local density approximation calculations on the other hand \cite{PhysRevB.77.046101}.
Based on the discussion of the Ru 3$d$ core level spectra \cite{PhysRevLett.93.126404}, Sr$_2$RuO$_4$ is more correlated than SrRuO$_3$. In the present work, we pointed out the correlation of LaRuO$_3$ and NdRuO$_3$ is even weaker than SrRuO$_3$ (Fig. \ref{fig:2} (b)). The 3 eV peak observed in LaRuO$_3$ and NdRuO$_3$ could thus yield another implications on the relationship between the correlation and 3 eV peak in Sr$_2$RuO$_4$.

The clear chemical shifts were observed in the O 1$s$ level in comparison with SrRuO$_3$ (Fig. \ref{fig:2} (c)). In particular, the binding energy of the O 1$s$ peak in LaRuO$_3$ and NdRuO$_3$ ($E_\text{B} \sim$ 529 eV) shifts upwards by $\sim$350 meV while the peak width is almost identical (Fig. \ref{fig:2} (d)). This shift and unchanged peak shape are qualitatively akin to hole-doped cuprates \cite{PhysRevLett.79.2101}. From the chemical point of view \cite{10.1063/1.4931996}, another O 1$s$ peak corresponding to the O$^{1-}$ ions, which could experimentally be observed as a broadened peak, should be present if the average valence of oxygen ions are different among these compounds. Nevertheless, we did not observe such effect, suggesting that the oxygen valence number in LaRuO$_3$ and NdRuO$_3$ is comparable with that of SrRuO$_3$ \cite{10.1063/1.4931996}. We thus associate the shift of the O 1$s$ peak with the different ruthenium valence states (i.e., trivalent and tetravalent) rather than the speculative change of the oxygen valence states. The binding energy of the shoulder peak is nearly identical among these three samples (Fig. \ref{fig:2} (d)). This sample-independent peak position also supports our argument that the shoulder peak is ascribed to surface contamination containing oxygen absorbed on the sample surface, also verified by our HAXPES data (Fig. \ref{fig:4} (g,h)).

\begin{figure}
  \centering
  \includegraphics[angle = 0, width = 0.48\textwidth, clip=true]{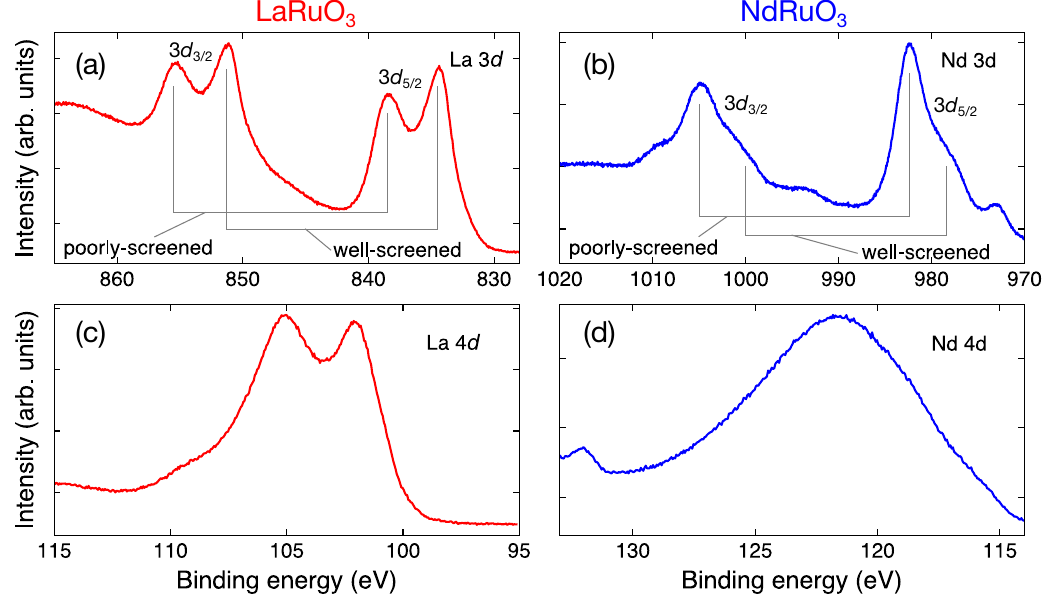}
  \caption{SXPES core level spectra of lanthanides. (a) La 3$d$. (b) Nd 3$d$. (c) La 4$d$. (d) Nd 4$d$. \label{fig:3}}
\end{figure}

Spectral shapes of 3$d$/4$d$ core levels in lanthanoid elements are also sensitive to their electronic states. 
From the chemical point of view, the valence number of La and Nd ions tends to be +3 in solids, but it is not trivial in the present case because ruthenium ions in oxides tend to be tetravalent. 
Figures \ref{fig:3} (a,b) display the 3$d$ core level spectra of La and Nd ions. In Fig. \ref{fig:3} (a), the peak splitting is notable besides the spin-orbit splitting between $3d_{5/2}$ and $3d_{3/2}$ peaks due to the Coulomb attraction between core holes and valence electrons peculiar to the photoemission process, also known as the final state effect.
As discussed in Ref. \cite{PhysRevLett.45.1597}, the well-screened and poorly-screened features are very clearly observed in both compounds. Even if the valence number is the same, the spectral shape could differ depending on the subtle balance between the charge transfer energy and core-hole potential. Nevertheless, the intensity ratio of these two features in LaRuO$_3$ and NdRuO$_3$ is comparable with that in La$_2$O$_3$ and Nd$_2$O$_3$, respectively. Our data thus imply that the trivalent La and Nd ions are surrounded by cations and anions in a similar fashion to La$_2$O$_3$ and Nd$_2$O$_3$ \cite{PhysRevB.32.6819}. The 4$d$ core level spectra of La and Nd (Fig. \ref{fig:3} (c,d)) can also be understood in the same manner \cite{Kotani1992Theory-of-}.
Overall, these SXPES spectral shapes are akin to those of trivalent La and Nd ions. Considering that the oxygen valence state is similar to the one in SrRuO$_3$ (Fig. \ref{fig:4} (d)), we conclude that the spectral shifts of Ru and O core levels in LaRuO$_3$ and NdRuO$_3$ are peculiar to the trivalent state of ruthenium ions, which were previously revealed by the Ru $K$-edge XAS \cite{2305.09201}.

In conclusion, we have investigated the core-level and valence band spectra of the LaRuO$_3$ and NdRuO$_3$ thin films by means of SXPES and HAXPES experiments. We found that the core level spectra are markedly different from those in Ru$^{4+}$ ions in SrRuO$_3$. We therefore claim that these spectra in LaRuO$_3$ and NdRuO$_3$ are peculiar to the trivalent ruthenium valence states.  This argument is supported by the core-level spectra of the cations (La and Nd) as well. We also found the 3 eV peak near the Fermi level, reminiscent of that observed in Sr$_2$RuO$_4$. This peak feature could yield insights into the origin of the 3 eV peak in  Sr$_2$RuO$_4$ in a fresh fashion. Although the pseudospin splitting near the Fermi level was not detected in the present study, it is highly desirable to investigate low-energy electronic states with high energy-resolution spectroscopy such as RIXS, which could potentially probe the excitation from the $j=1/2$ doublet to the $j=3/2$ quartet in a similar fashion to other 4$d^5$ systems, e.g., $\alpha$-RuCl$_3$ \cite{Lebert_2020,Suzuki2021Proximate-} in part because LaRuO$_3$ and NdRuO$_3$ films are rare materials accommodating 4$d^5$ electrons in the square lattice.


We thank Arata Tanaka for informative discussions. The present work is supported by JSPS Grant-in-Aid for Scientific Research on Innovative Areas “Quantum Liquid Crystals” No. 19H05824, JSPS Grants-in-Aid for Scientific Research (S) No. JP22H04958, JSPS Grant-in-Aid for Early-Career Scientists No. JP20K15168, The Mitsubishi Foundation, The Izumi Science and Technology Foundation, and The Tokuyama Science Foundation. The HAXPES measurements were performed using the beamline BL24XU at SPring-8 (Proposal Nos. 2022A3231 and 2022B3231). We thank Kento Takenaka and Koji Takahara for their technical supports during the HAXPES measurements.

The data that support the findings of this study are available from the corresponding author upon reasonable request.

\section*{References}

\bibliography{papers.bib}

\end{document}